# Bat Optimized Watershed based Segmentation of Lamina Cribrosa

**Abhisha Mano, Assistant Professor, ECE department, Rajas International Institute of Technology for Women, Nagercoil.**

*Abstract*: The segmentation of Lamina Cribrosa(LC) is a challenging task to detect the glaucomatous damage. In this paper a new method of segmenting the LC using bat optimized Watershed segmentation is done. By using wavelet transform LC structures are decomposed. Then, the decomposed image is optimized using Bat algorithm and by applying histogram equalization the optimized image is normalized. Watershed algorithm is used to segment the Lamina Cribrosa from its outer layer. Using some parameters like PSNR, MSE, F-Measure, rand index, sensitivity, specificity, SSIM and accuracy, the performance of the proposed system is calculated. The results show that the proposed method provides higher accuracy of 99.29%.

## 1.INTRODUCTION

The Lamina Cribrosa (LC) is a mesh featured arrangement in the later eye.LC is the location from where retinal ganglion cell (RGC), which is the axon from retina to the brain with visual data pass through the sclera. It has a structure of mesh shaped beams, having a collagen-rich connective tissue [10]. Glaucoma is an eye syndrome, initiated due to intraocular pressure which causes injury to optic nerve. This may cause eye blindness and information to the brain is also lost. The disease cannot be diagnosed at an early stage, and hence blindness occurs easily. Lamina Cribrosa helps to find out the glaucoma stage [3]. Imaging methods like Heidelberg retinal tomography and optical coherence tomography (OCT) helps to know the tissue structure of LC [7], [8]. This measures the thickness of retinal nerve fiber layer [9],[10] and loss that occurs in retinal epithelial cells [8], [11], [12] and other retinal layers [13], [8],[15].

An approach which is used to improve the contrast of the microstructure in micro-computed tomography imaging. A Frangi's filter is used to segment and define the orientation of each and every LC tissue, from micro-CT and SHG microscopy. An approach is used for the segmentation of the anterior Lamina Cribrosa surface. A Markov random field using shape parameters, estimated from a Metropolis-Hastings algorithm is used for segmentation. Using a approach, from scanning electron microscopy the fine structure of the Lamina Cribrosa is observed. The lamina at sclera contain larger pores .Since the laminar zones are the sites of glaucoma damage. A framework for defining the optic nerve head (ONH) is proposed. Our approach deals with tissues of the ONH at low Intraocular Pressure(IOP)[2][3].A method for segmenting the anterior LC in the images through multiphoton microscopy. A 4-D collaborative filtering is used to remove the noise. A wavelet multiresolution analysis used to enhance the structures. A morphological area opening is used to remove 3-D regions in the binarized images [14] [6][4]. Superpixel segmentation by enhancing the contrast of retinal images is presented in paper [15]. A method of segmenting mammogram and DNA fragments are shown in paper[16][17].

Earlier segmentation algorithms have shown good results only for OCT images. A novel and efficient method is proposed in this paper for segmenting LC from MPM images. In this work multiscale wavelet decomposition with adaptive scale selection is used. This algorithm handle segmentation of highly inhomogeneous LC beams.

## 2.PROPOSED METHODOLOGY

The input image is taken from the database and given to the noise removal section, where the noise is removed using a filter. The flow diagram of this proposed method is shown in fig 1.

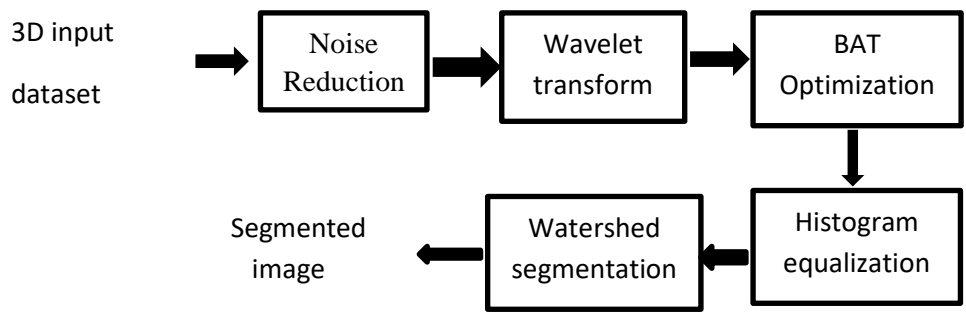

Fig. 1 Block diagram of the Proposed System

### a) Wavelet Transform:

Undecimated wavelet transform is applied on both down sampling and upsampling in case of forward wavelet transform and inverse wavelet transform respectively. The IUWT is a method to decompose an image into varying scales. Fig 2 and Fig 3 shows the input image and the wavelet transformed image respectively.

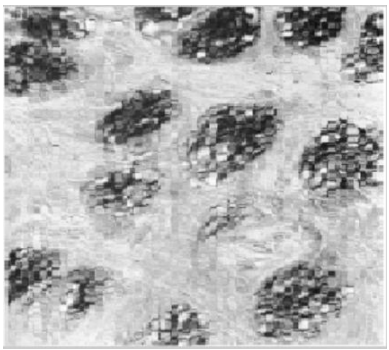 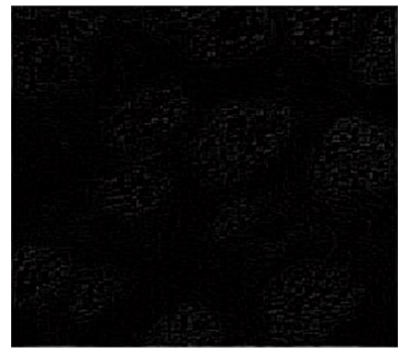

Fig 2 Input image  Fig.3 Wavelet transformed image

### b) BAT Algorithm:

Bat Algorithm is a metaheuristics method of solving optimization problems. Bat Algorithm is based on the activities and behaviour of bats. Each and every bat signifies only one solution search space. There are two real valued n-dimensional vectors. One vector represents, position of a bat. The other represents velocity of bats. Generally, position vector and velocity vector are initialized in a random manner at the starting of the algorithm. At each and every iteration fitness value is calculated for bats. A new velocity vector is calculated. Next, position of every bat is updated depending on its velocity vector. This process is repeated till convergence is reached. Fig 4 shows the convergence curve of the bat algorithm with iteration in the x-axis and average best value in the y-axis having the iteration as 500. Here best pixel intensity value is obtained.

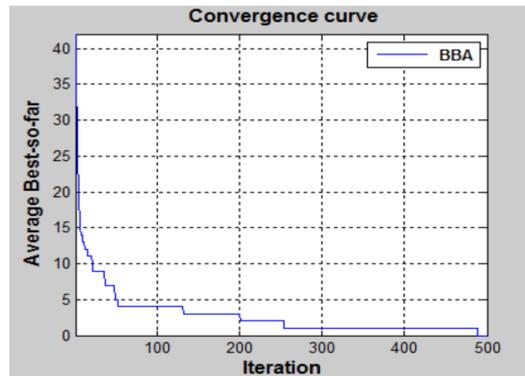

Fig. 4 Convergence curve for Bat Algorithm

**c)Histogram Equalization:**

Histogram equalization enhance contrast by adjusting the intensities. This technique provides good quality images. After applying histogram equalization, noise hidden in the image can be obtained. Here in addition to this a region is selected and it is cropped.

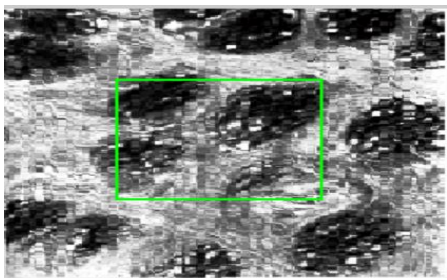 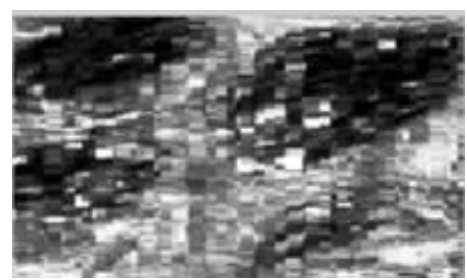

Fig.5 Histogram equalized image                Fig. 6 Cropped image

**d)Watershed Segmentation:**

In watershed segmentation the edges of the high intensity pixels are first detected by using edge detection, then by using the gradient magnitude, the segmentation process is done. Watershed Segmentation uses the Gradient Magnitude as the Segmentation Function. Fig 7 shows the segmented output.

To mark the boundaries of the segmented image, area localization is done. The red color marking in the segmented image shows the lamina cribrosa area of the image. The remaining portion shows the hole region in the input image.

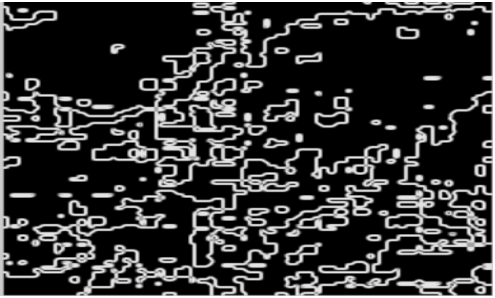 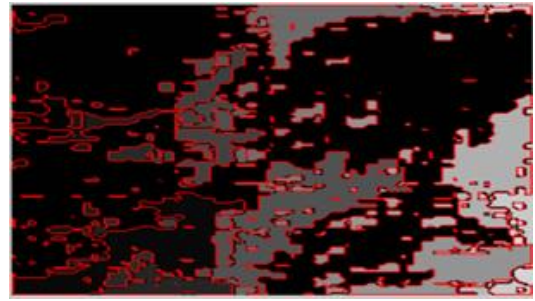

(a) (b)

Fig.7 Segmented output of the Lamina Cribrosa for a specified region

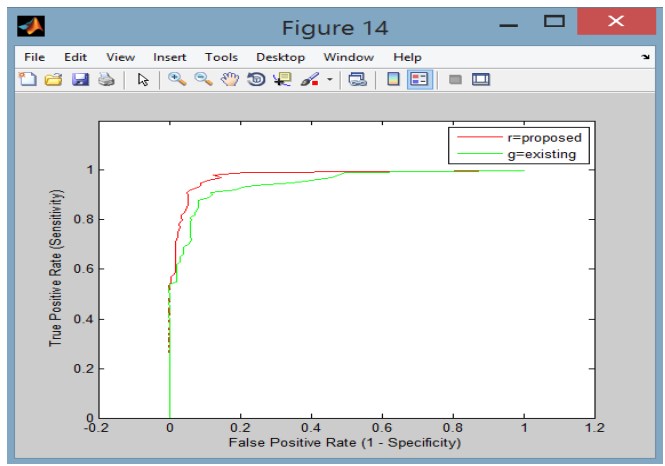

Fig 8  ROC plot

Some parameters like PSNR, MSE, F-Measure, rand index, sensitivity, specificity, SSIM and accuracy are found.

Table 1. Performance Measures

| Performance measures | |
|---|---|
| PSNR | 58.65 |
| MSE | 0.08 |
| F-Measure | 1.95 |
| Rand Index | 0.98 |
| Sensitivity | 100 |
| Specificity | 98.59 |
| SSIM | 99.29 |
| Accuracy | 99.291 |

The performance of the proposed method is evaluated with graph cut method without optimization and ROC is plotted. Fig 8 shows the ROC chart of the proposed method. Table 1 shows the performance measures which are done in this approach. From this it is understood that the proposed method provides an accuracy of 99.29% .

## 3. CONCLUSION

In this paper, a new approach which is based on the BAT optimized Watershed segmentation was done to segment LC. The experimental results prove that the proposed method provides an accuracy of 99.29%.

**REFERENCES**

1. Sundaresh Ram, *Member,* Forest Danford, Stephen Howerton, Jeffrey J. Rodr´ıguez, *Senior Member,* and Jonathan P. Vande Geest, Three-Dimensional Segmentation of the Ex-Vivo Anterior Lamina Cribrosa from Second-Harmonic Imaging Microscopy.
2. R. N. Weinreb, T. Aung, and F. A. Medeiros, "The pathophysiology and treatment of glaucoma: a review," *Jama*, vol. 311, no. 18, pp. 1901–1911, May 2014.
3. H. A. Quigley and A. T. Broman, "The number of people with glaucoma worldwide in 2010 and 2020," *Brit. J. Ophthalmol.*, vol. 90, no. 3, pp. 262–267, 2006
4. C. Kervrann, C. O. S. Sorzano, S. Acton, J.-C. Olivo-Marin, and M. Unser, "A guided tour of selected image processing and analysis methods for fluorescence and electron microscopy," *IEEE J. Select. Topics Signal Process.*, vol. 10, no. 1, pp. 6–30, Feb. 2016.
5. B. Zhang, M. J. . Fadili, and J. L. Starck, "Wavelets, ridgelets, and curvelets for Poisson noise removal," *IEEE Trans. Image Process.*, vol. 17, no. 7, pp. 1093–1108, Jul. 2008.
6. F. Luisier, T. Blu, and M. Unser, "Image denoising in mixed Poisson-Gaussian noise," *IEEE Trans. Image Process.*, vol. 20, no. 3, pp. 696–708, Mar. 2011.
7. I. I. Bussel, G. Wollstein, and J. S. Schuman, "OCT for glaucoma diagnosis, screening and detection of glaucoma progression," *Brit. J.Ophthal.*, pp. bjophthalmol–2013, Dec. 2013.
8. E. A. Gibson, O. Masihzadeh, T. C. Lei, D. A. Ammar, and M. Y. Kahook, "Multiphoton microscopy for ophthalmic imaging," *J. Ophthalmol.*, vol. 2011, p. 870879, 2011.
9. K. A. Townsend, G. Wollstein, and J. S. Schuman, "Imaging of the retinal nerve fibre layer for glaucoma," *Brit. J. Ophthalmol.*, vol. 93,no. 2, pp. 139–143, Feb. 2009.
10. Z. Nadler, B. Wang, G. Wollstein, J. E. Nevins, H. Ishikawa, L. Kagemann, I. A. Sigal, R. D. Ferguson, D. X. Hammer, I. Grulkowski, J. J. Liu, M. F. Kraus, C. D. Lu, J. Hornegger, J. G. Fujimoto, and J. S.Schuman, "Automated lamina cribrosa microstructural segmentation inoptical coherence tomography scans of healthy and galucomatous eyes," *Biomed. Opt. Express*, vol. 4, no. 11, pp. 2596–2608, 2013.
11. J. M. Bueno, A. Giakoumaki, E. J. Gualda, F. Schaeffel, and P. Artal, "Analysis of the chicken retina with an adaptive optics multiphoton microscope," *Biomed. Opt. Express*, vol. 2, no. 6, pp. 1637–1648, Jun. 2011.
12. O. Masihzadeh, T. C. Lei, D. A. Ammar, M. Y. . Kohook, and E. A.Gibson, "A multiphoton microscope platform for imaging the mouseeye," *Molecular Vision*, vol. 18, pp. 1840–1848, Jul. 2011.
13. O. Tan, G. Li, A. T. Lu, R. Varma, and D. Huang, "Mapping of macular substructures with optical coherence tomography for glaucoma diagnosis," *Ophthalmology*, vol. 115, no. 6, pp. 949–956, Jun. 2008.

14. M. Maggioni, V. Katkovnik, K. Egiazarian, and A. Foi, "Nonlocal trans-form domain filter for volumetric data denoising and reconstruction," *IEEE Trans. Image Process.*, vol. 22, no. 1, pp. 119–133, Jan. 2013.
15. Abhisha Mano, "Contrast Enhanced Superpixel Based Segmentation Of Retinal Images" TechRxiv. Preprint. https://doi.org/10.36227/techrxiv.12254240.v1. 2020
16. S. Anand, B.Lakshmanan, J. Murugachandravel, K. Valarmathi, Abhisha Mano, N. Kavitha," Wavelet-Based Automated DNA Sizing of Fragments through AFM Image Processing", International Journal of Engineering and Advanced Technology (IJEAT), Volume-8 Issue-5, June 2019.
17. S. Anand, J. Murugachandravel, K.Valarmathi, Abhisha Mano, N. Kavitha," Contrast Enhancement of Mammograms and Microcalcification Detection", International Journal of Recent Technology and Engineering (IJRTE), Volume-8 Issue-4, November 2019.